\documentclass{article}
\pdfoutput=1
\usepackage{graphicx} 
\begin{document}
\title{A simple explanation of OPERA results without strange physics}

\author{G. Henri \\
        UJF-Grenoble 1 / CNRS-INSU\\
        Institut de Plan\'etologie et d'Astrophysique de Grenoble (IPAG)\\
         UMR 5274, Grenoble, F-38041, France\\
        E-mail: Gilles.Henri@obs.ujf-grenoble.fr\\}
 \maketitle

\abstract
{We show that OPERA recent results  \cite{Adam11} showing an apparent superluminal velocity of muonic neutrinos can find a very simple explanation without any measurement error or any strange physics. Namely, it is enough that the beam composition varies during the leading and the trailing edges to explain an apparent time shift in the detected neutrinos.  The order of magnitude of the shift will be the relative variation of the average cross-section times the rising/decaying time, and even a modest change in the composition of the beam could produce the observed effect.

\section{Introduction}
The recent claim by the OPERA collaboration of an apparent superluminal motion of muonic neutrinos has raised a considerable interest in the particle physics community. The collaboration claims to have detected an unexplained residual time shift of 60 ns of the neutrino signal with respect to the time of flight (TOF) corresponding to the light travel time, computed thanks to an accurately determined distance between the source at CERN and the target at Gran Sasso (730.085 km). The apparent negative delay is around 60 ns, corresponding to a distance shorter than expected by around 20 m. The collaboration claims a much more accurate determination of the distance, ruling out a possible error in its determination. Various explanations have been proposed ranging from bad determination of error bars, retracted shortly after (http://johncostella.webs.com/neutrino-blunder.pdf), problems with synchronization of clocks \cite{cont11},  or trying to explain the result by really tachyonic and unconventional physics of neutrinos (\cite{Gaccia11}-\cite{Ciuffol11}). One of the most striking feature of the detection is the lack of significative dependence on average energy ( \cite{Adam11}). By splitting the sample in two energy bins with the same statistics, the average energy of each subsample differ by a factor around 3 (13.9 GeV and 42.9 GeV respectively). However, no significative difference in time delays are obtained, although a marginally significative difference has been measured ($53.1 \pm 18.8 \rm{ (stat)} \pm 7.4 \rm{ (sys)}$ ns vs $67.1 \pm 18.8 (\rm{stat}) \pm 7.4\rm{( sys)}$ ns ) . This is almost incompatible with a linear dependence with energy, although the upper limits derived from electronic neutrinos detected from the supernova SN1987A would imply a factor 1000 lower in the velocity shift, with an energy also around a factor 1000 lower (10 MeV instead of 30 GeV). Although one can argue a different physics for electronic muonic neutrinos, or invoke the interaction with matter, and so on, it is uncomfortable to invoke unnatural behaviors (\cite{Fargion11}, \cite{Gonzalez11}). Superluminal neutrinos should also experience a strong attenuation due to kinematically allowed electron-positron pair production \cite{Cohen11}. We propose here a possible explanation which involves no measurement error and no unconventional physics. Namely, we notice that it is enough that the energy distribution of the beam changes during the leading and trailing edges , so that the probability detection is itself a function of time. This will modify the shape of the transition periods, that are dominant in the statistical comparison of initial and final beams. We show that the main effect is a first order time shift, that can be of either sign depending on the variation of the average cross section. We illustrate this effect by a simple example and show that a modest decrease of the average energy of the neutrino beam could produce the observed apparent superluminal shift. 

\section{Effect of a varying effective cross section}

To illustrate the proposed effect, we assume that the intensity of the beam is a "logistic" function (Verhulst 1845),  growing from zero to one in properly normalized units. We normalize the time $t$ to the characteristic rising time $t_r$ by setting $\tau = t/tr_r$, the origin of time being the "nominal" light travel time at the expected velocity $v  \simeq c$. The resulting intensity will then read \\
\begin{equation}
I(\tau) = \frac{I_0}{1+e^{-\tau}}
\label{Iinit}
\end{equation}

Now we assume that the average cross section, governing the detection probability, is itself varying from an initial value $\sigma_{i}$ to a final value $\sigma_f=\sigma_i (1+\epsilon) $, where $\epsilon = \Delta \sigma /\sigma$ is the relative variation of the averaged cross section during the rising phase. We assume the same logistic dependence of $\sigma$

\begin{equation}
\sigma(\tau)= \sigma_i (1+\frac{\epsilon}{1+e^{-\tau}})
\end{equation}

The apparent  flux measured by the detection device will be 

\begin{equation}
F_{app} \propto I(\tau)\sigma(\tau) = \frac{ I_0 \sigma_i }{1+e^{-\tau}}(1+\frac{\epsilon}{1+e^{-\tau}})
\label{Ifin}
\end{equation}

However, fitting this profile with the initial profile (\ref{Iinit}) shifted in time means that we try to fit the function (\ref{Ifin}) by a trial function 

\begin{equation}
I(\tau) = A. \frac{I_0}{1+e^{-(\tau-\Delta \tau)}}
\label{trial}
\end{equation}
$A$ and $\Delta \tau$ being the free parameters of the fit. Normalization of the asymptotic plateau provides $A = 1+ \epsilon$, so the minimization procedure will actually minimize the residuals 
\begin{equation}
R = \Sigma_j \left[  \frac{1 }{1+e^{-\tau_j}}(1+\frac{\epsilon}{1+e^{-\tau_j}}) -  \frac{1+\epsilon}{1+e^{-(\tau_j-\Delta \tau)}}\right]^2
\end{equation}

As the shape of the rising curve will be somewhat different from the initial one, this will generally lead to a non-zero $\Delta \tau$, although none of the particle is actually travelling faster than light. 
A first order approximation of $\Delta \tau$ as a function of $\epsilon$ can be found by remarking that the fit should approximately reproduce the half rising time when the intensity reaches half of the maximal one.  Setting $I_{app}(\Delta \tau) = I_{max}/2 = I_0 (1+\epsilon)/2 $, one gets 
\begin{equation}
\frac{1+\epsilon/(1+e^{-\Delta \tau })}{1+e^{-\Delta \tau}}= \frac{1+\epsilon}{2}
\end{equation}

which, for small $\Delta \tau$, leads eventually to $\Delta  \tau \simeq \epsilon$. {\em The apparent shift in TOF is thus of the order of the relative variation of the cross section times the rising time of the bunch}.

Note that an apparent advance of the rising edge can be obtained with {\em negative} values of $\epsilon$, that is average cross section decreasing with time.  We confirm this estimate by trying a numerical optimization of the function (\ref{Ifin}) by the trial function (\ref{trial}). The result is displayed Fig. 1 for $ \epsilon = - 0.2$, the numerical optimum being $\Delta \tau = -0.24$. As can be seen , the resulting fit is almost indistinguishable from the original curve, broadening effects being at second order only. Thus only the first order time shift can be detected, giving a false impression of travelling faster than light, although none of the individual particle does.
\begin{figure}
\begin{center}
\includegraphics[width=0.9\linewidth]{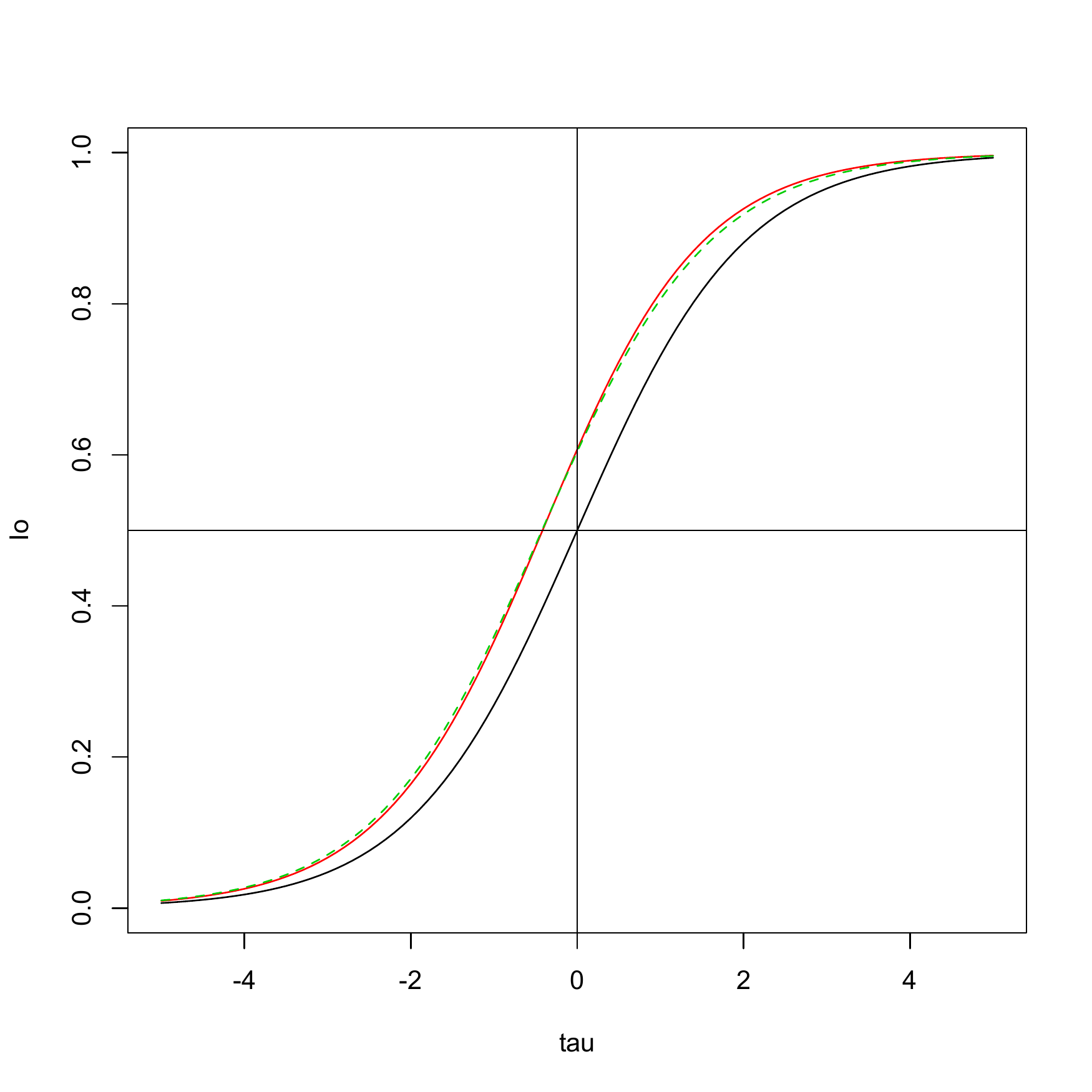}
\caption{Simulation of the effect of variation of the averaged cross section on the apparent leading edge of a signal. Black solid line : "true" intensity (shifted by the light TOF ) at the nominal time. Red solid line : apparent intensity with a 20 percent decrease of the cross section, after renormalization (Eq. (3)). Green dashed line : best fit of the second curve by a shifted "true" intensity, corresponding to a (negative) delay $\Delta \tau = -0.24$}
\label{fig1}
\end{center}
\end{figure}

\section{Discussion}
Turning to actual OPERA results, we see (Fig. (12) of the collaboration paper) that the leading edges have characteristic timescales about 500 ns, whereas the measured time shift is only around 50 ns. Thus a mere variation of 10 \% of the averaged cross section could explained the observed shift. As the shift is of the same order of magnitude at the two edges, by symmetry, the cross section should decrease by around 10 \% both at leading and trailing edge. The muonic neutrino interaction cross section increases with energy, so it would mean that the average energy is decreasing during both phases. It is beyond the scope of this paper to study the energy dependence of the neutrino beam during its rising and decaying phase, but it would be valuable for experimentalists to ascertain this parameter. A crucial test of this explanation is that the observed shift is not dependent on the travelled distance, being only a mere consequence of the pulse shape and composition. This would naturally explain the strange (and constant) value of $v-c \ll c$, which doesn't fit easily in any expected tachyonic model. Also one could have a possible dependence on average energy, since nothing tells that the relative variation of cross section is the same over any energy distribution, but it is not expected that it would scale linearly with energy. However, the apparent delay would be a constant fraction of the rising time, so a dependence of this delay with the shape of the beam would be a clear evidence for the validity of the proposed mechanism.

\section{Conclusion}
We propose a possible explanation of OPERA results without invoking any violation of Lorentz invariance and without any experimental errors. The observed apparent time shift could be entirely accounted for by a small variation of the averaged interaction cross section, being itself due to a variation of the energy spectrum. A mere 10 \% variation is enough to account for the observed value, a leading pulse being obtained by a decreasing average cross section with time, i.e. for neutrinos, a decreasing average energy with time, both in the leading edge and in the trailing edge.  A crucial prediction of this model would be that the time shift is independent of the distance, but should depend on the pulse shape.


\begin{thebibliography}{}

\bibitem{Adam11}
T.~Adam, {\em et al.} (OPERA collaboration), 2011, hep-ex/1109.4897 
\bibitem{cont11}
C.R. Contaldi , 2011, hep-ph/1109.6160
\bibitem{Gaccia11}
G. Gacciapaglia A. Deandrea, and L. Panizzi , 2011, hep-ph/1109.4980
\bibitem{Amelino11}
G. Amelino-Camelia {\em et al.}.  , 2011, hep-ph/1109.5172
\bibitem{Tamburini11}
F. Tamburini  and M. Laveder , 2011, hep-ph/1109.5445
\bibitem{Gubser11}
S. S. Gubser , 2011, hep-th/1109.5687
\bibitem{Oiko11}
V.K.Oikonomou, 2011, hep-th/1109.6170
\bibitem{Lorio11}
L. Lorio, 2011, gr-qc/1109.6249
\bibitem{Hannes11}
S. Hannestad and M. Sloth , 2011, hep-ph/1109.6282
\bibitem{Kehag11}
A. Kehagias, 2011, hep-ph/1109.6312
\bibitem{Gardner11}
S. Gardner, 2011, hep-ph/1109.6520
\bibitem{Klink11}
F.R. Klinkhamer and G.E. Volovik, 2011, hep-ph/1109.6624
\bibitem{Matone11}
M. Matone, 2011, hep-ph/1109.6631
\bibitem{Ciuffol11}
E. Ciuffoli {\em et al.}, 2011, hep-ph/1109.6520
\bibitem{Cohen11}
A. G. Cohen and S. L. Glashow, 2011, hep-ph/1109.6562
\bibitem{Giudice11}
G.F. Giudice et al. , 2011, hep-ph/1109.5682
\bibitem{Fargion11}
D. Fargion, 2011, astro-ph.HE/1109.5368
\bibitem{Gonzalez11}
L. Gonzales-Mestres, 2011, physics.gen-ph/1109.6630

  \end{thebibliography}
\end{document}